\documentclass[twocolumn,prl]{revtex4}
\usepackage{times,graphicx,amsmath}
\usepackage{amssymb}
\usepackage{epsfig}

\newcommand{\be}{\begin{equation}}
\newcommand{\ee}{\end{equation}}
\newcommand{\bea}{\begin{eqnarray}}
\newcommand{\eea}{\end{eqnarray}}
\newcommand{\avg}[1]{\langle{#1}\rangle}

\newcommand{\bra}[1]{\langle{#1}|}
\newcommand{\braket}[2]{\langle{#1}|{#2}\rangle}
\newcommand{\ket}[1]{|{#1}\rangle}
\begin{document}

\title{Dynamic instability in a phenomenological model of correlated assets}
%{How much investment can financial markets absorb?}
\author{Giacomo Raffaelli \& Matteo Marsili}

\affiliation{INFM-SISSA, via Beirut 2-4, Trieste I-34014, Italy and\\
Abdus Salam International Centre for Theoretical Physics\\
Strada Costiera 11, 34014 Trieste, Italy}

\begin{abstract}
We show that financial correlations exhibit a non-trivial dynamic
behavior. We introduce a simple phenomenological model of a
multi-asset financial market, which takes into account the impact
of portfolio investment on price dynamics. This captures the fact
that correlations determine the optimal portfolio but are affected
by investment based on it. We show that such a feedback on
correlations gives rise to an instability when the volume of
investment exceeds a critical value. Close to the critical point
the model exhibits dynamical correlations very similar to those
observed in real markets. Maximum likelihood estimates of the
model's parameter for empirical data indeed confirm this
conclusion, thus suggesting that real markets operate close to a
dynamically unstable point.
\end{abstract}
\maketitle

Financial markets -- as prototypical examples of the collective
effects of human interaction -- have recently attracted the
attention of many physicists. This is because, in spite of their
internal complications, their aggregate behavior exhibits
surprising regularities which can be cast in the form of simple,
yet non-trivial, statistical
laws\cite{JPB-book,MantegnaStanley,Daco}, reminiscent of the
scaling laws obeyed by anomalous fluctuations in critical
phenomena. Such a suggestive indication has been put on even firmer basis by
recent research on the statistical physics approach to interacting
agent models\cite{mercatino,Lux,contbouchaud,levy,MG}. This has
shown that quite realistic market behavior can indeed be generated
by the internal dynamics generated by traders' interaction.

The theoretical approach has, thus far, mostly concentrated on
single asset models, whereas empirical analysis has shown that
ensembles of assets exhibit rich and non-trivial statistical
properties, whose relations with random matrix theory 
\cite{jpb-rmt,gopicorr}, complex networks \cite{kertesz,gcalda} and
multi-scaling \cite{Eisler} have attracted the interest of 
physicists. 
The central object of study is the covariance matrix
of asset returns (at the daily scale in most cases). 
The bulk of
its eigenvalue distribution is dominated by noise and described
very well by random matrix theory \cite{jpb-rmt}. The few large
eigenvalues which leak out of the noise background contain
significant information about market's structure. The taxonomy
built with different methods \cite{mantegna,gopicorr,states}
from financial correlations alone bear remarkable similarity with
a classification in economic sectors. This agrees with the
expectation that companies engaged in similar economic activities
are subject to the same ``factors'', e.g. fluctuations in prices
or demands of common inputs or outputs. Besides their structure,
market correlations also exhibit a highly non-trivial dynamics:
Correlations ``build up'' as the sampling time horizon on which returns are
measured increases (Epps effect) and saturate for returns on the 
scale of some days \cite{Drozdztau}. Furthermore, these correlations 
are persistent over time \cite{kertesz} and they
follow recurrent patterns \cite{states}. 

\begin{figure}[ht]
  \centerline
      {
        \hbox
            {
              \psfig{figure=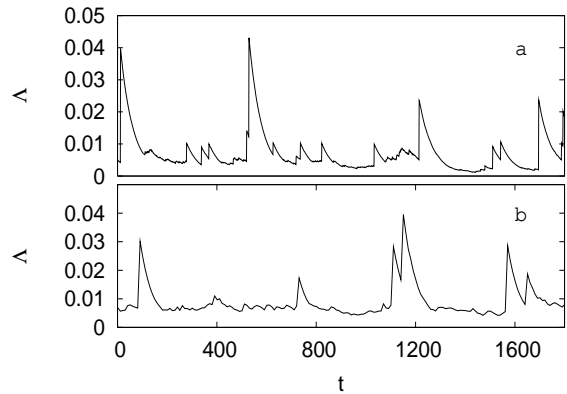,height=6cm,width=8cm}
            }
      }
      \caption{Maximum eigenvalue of the correlation matrix
        as a function of time for $\tau=50$. a) Toronto Stock exchange \cite{data}.
        Here the correlation matrix is obtained using Eqs. (\ref{ccc}) and (\ref{cc}) with $\ket{\delta x_t}$ taken from historical data.
        b) simulation of Eq.(\ref{eq1}) with
        $N=20$, $R=5 \,10^{-4}$, $B=10^{-3}$,$\Delta = 1$,$\epsilon=10^{-1}$,
    $W=0.245$.
        Components of $\ket{b}$ where generated uniformly in the interval $[0,2\cdot 10^{-3}]$, resulting in $W^*\approx 0.25$.
      }

     \label{Lambdatime}
\end{figure}

\begin{figure}[ht]
  \centerline
      {
        \hbox
            {
              \psfig{figure=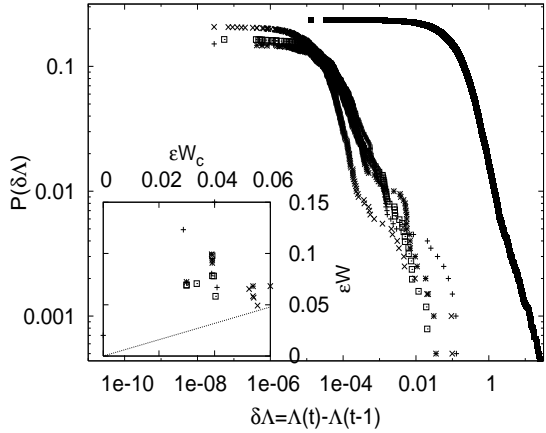,height=6cm,width=8cm}
            }
      }
      \caption{Cumulative distribution of the day-to day change in the maximum
eigenvalue for different indices: DAX ($+$), TSX($\star$), DOW($\square$), ASX($\times$).
Also shown is the same distribution for a numerical simulation of the model with $N=20$,
$\epsilon=0.1$, $R=1$, $W=14.7$, $\Delta=1$,$B=10^{-2}$,$\tau=100$. In the inset we show
the results of the fitting procedure on these indices (same symbols).}
      \label{PL-wwc}
\end{figure}

In what follows we shall mostly concentrate on the dynamics of the
largest eigenvalue of the correlation matrix. The corresponding
{\em market mode} \cite{jpb-rmt} describes the co-movement of
stocks and it accounts for a significant fraction of the
correlations\cite{nota_ec}. Fig. \ref{Lambdatime}a shows the time
dependence of the largest eigenvalue of the (exponentially
averaged) correlation matrix of daily returns for Toronto Stock
Exchange \cite{data}. Similar behavior has been reported earlier
\cite{Drozdz} for different markets. Fig. \ref{PL-wwc} shows that
fluctuations in the largest eigenvalue are broadly distributed,
suggesting that Fig. \ref{Lambdatime}a can hardly be explained
entirely as the effect of few external shocks.

This leads us to formulate the hypothesis that such non-trivial
behavior arises as a consequence of the internal market dynamics. One
of the key functions of financial markets is indeed that of allowing
companies to ``trade'' their risk for return, by spreading it across
financial investors. Investors on their side, diversify (i.e. spread)
their strategies across stocks so as to minimize risk, as postulated
by portfolio optimization theory \cite{elton}. The
efficiency of portfolio optimization depends on the cross
correlations among the stocks the financial market is composed of.
The optimal portfolio is computed under the {\em price taking} assumption
that investment does not affect the market. While this is reasonable for
the single investor, the effect of many investors following this same strategy
can be sizeable. If financial trading activity resulting from portfolio
optimization strategies have an impact on prices' dynamics, it will also
affect the correlations which these strategies exploit.
Hence financial correlations enter into a feedback loop
because they determine in part those trading strategies which
contribute to the price dynamics, i.e. to the financial correlations
themselves. This feedback is somewhat implicit in the Capital Asset
Pricing Model (CAPM), which concludes that since all traders invest
according to the optimal portfolio, the market is well approximated by
a one factor model \cite{elton} (see however \cite{gcalda}). While this
explains why the largest eigenvalue of the correlation matrix is so
well separated from the other ones, CAPM relies on rational
expectation equilibrium arguments, and it does not address dynamical
effects such as those of Fig. \ref{Lambdatime}a.

This Letter discusses a general phenomenological approach, in the spirit 
of Landau's theory of critical phenomena \cite{Landau}, which shows 
that a non-trivial dynamics of correlations can indeed
result from the internal dynamics due to trading on 
optimal portfolio strategies. The model predicts a
dynamical instability if the investment volume $W$ exceeds a
critical value. Not only we find very realistic dynamics of
correlations close to the critical point (see Fig.
\ref{Lambdatime}b) but maximum likelihood parameter estimation
from real data suggest that markets are indeed close to the
instability. Phenomenological models are particularly suited for modeling complex systems,
such as a financial market, were a bottom-up (microscopic) approach inevitably 
implies dealing with many complications and introducing {\em ad hoc} 
assumptions \cite{note}. For the ease of exposition, we shall first  
introduce a minimal model which captures the interaction among assets induced 
by portfolio investment. 
Later we will show that this model contains the lowerst order
terms in a general expansion of the dynamics and that all the 
terms beyond these are irrelevant as far as the main 
conclusions are concerned. A further reason for focusing on the simplest model
is that it will make the comparison with empirical data easier.

Let us consider a set of $N$ assets. We denote by $\ket{x}$ the vector of log-prices 
and use bra-ket notation \cite{bra-ket}.
We focus on daily time-scale and assume that $\ket{x_t}$
undergoes the dynamics

\begin{equation}\label{eq1}
\ket{x_{t+1}}=\ket{x_t}+\ket{\beta_t}+\xi_t\ket{z_t}.
\end{equation}
where $\ket{\beta _t}$
is the vector of {\em bare} returns, 
which describes all {\em external} ``forces'' which drive the prices, including
economic processes.
This is assumed to be a Gaussian random vector with

\begin{equation}\label{bare}
  E[\ket{\beta_t}]=\ket{b},~~~~~E[\ket{\beta_t}\bra{\beta(t')}]=
\ket{b}\bra{b}+\hat B\delta_{t,t'}
\end{equation}
$\ket{b}$ and $\hat B$ will be considered as
parameters in what follows.

The last term of Eq. (\ref{eq1}) describes the impact of
portfolio investment on the price dynamics: $\xi_t$ is
an independent Gaussian variable with mean
$\epsilon$ and variance $\Delta$ and the vector $\ket{z_t}$ is
the optimal portfolio with fixed return $R$ and total 
wealth $\braket{z}{1} = W$.
In other words, $\ket{z_t}$ is the solution of

\begin{equation}\label{eq2}
  \min_{\ket{z},\nu,\sigma}\left[\frac{1}{2}\bra{z}\hat C_t\ket{z}-
\nu\left(\braket{z}{r_t}-R\right)
-\sigma\left(\braket{z}{1}-W\right)
  \right]
\end{equation}
where $\hat C_t$ is the correlation matrix at time $t$. Both the expected
returns $\ket{r_t}$ and the correlation matrix $\hat C_t$, which
enter Eq. (\ref{eq2}), are computed from
historical data over a characteristic time $\tau$:

\begin{eqnarray}\label{ccc}
\ket{r_t}&=&(e^{\frac{1}{\tau}}-1)\sum_{t'<t}e^{-\frac{t-t'}{\tau}}
\ket{\delta
  x_t}\\
\hat
C_{t}&=&(e^{\frac{1}{\tau}}-1)\sum_{t'<t}e^{-\frac{t-t'}{\tau}}
\ket{\delta
  x_t}\bra{\delta x_t}-\ket{r_t}\bra{r_t}\label{cc}
\end{eqnarray}
where $\ket{\delta x_t}\equiv\ket{x_{t+1}}-\ket{x_t}$ \cite{noteavg}. 
This makes the set of equations above a self-contained dynamical stochastic system. 
In a nutshell, it describes how the economic {\em bare} correlated fluctuations 
$\ket{\beta_t}$ are {\em dressed} by the trading activity due to investment in 
optimal portfolio strategies.
$R,~W,~\Delta$ and $\tau$ are {\em phenomenological} parameters reflecting the 
portfolio composition which dominates trading activity, not necessarily those 
of a representative investor. 
In particular, note that $\ket{z_t}$ is a quantity, not a percentage as
in portfolio theory \cite{elton}. Indeed, the impact of investment on prices depends
on the volume of transactions. The parameter $W$, which can be taken as a proxy for 
the volume of trading on portfolio strategies, will play a crucial role in what 
follows.

These parameters are assumed to be constant, meaning that 
portfolio policies change over time-scales much longer than $\tau$.
Portfolio theory is, in principle, based on expected returns and covariances. 
We implicitely assume that historical data can be used
as a proxy for expected correlations and returns. Eqs. (\ref{eq1}--\ref{cc}) also assume 
that portfolio investment is dominated by a single time scale $\tau$. Later we shall
argue that a generic distribution of time scales would not change the main results. 
Finally, Eq. (\ref{eq1}) assumes a linear price impact and gaussian {\em bare} returns. 
Both assumptions may be questionable, specially at high frequency \cite{high_freq_impact,impact}.
We shall see, however, that non-trivial dynamics and statistics (including a fat tailed 
distribution of returns) arises even in such a simplified setting,
thus suggesting that the specific market mechanism and the statistics of {\em bare} 
returns are unessential ingredients.

Numerical simulations of the model show a very interesting
behavior. In Fig \ref{Lambdatime}b we plot the temporal evolution
of the maximum eigenvalue of the correlation matrix for a
particular choice of parameters (see later). The dynamics is
highly non-trivial, with the appearance of instabilities
resembling those observed for real markets (Fig.
\ref{Lambdatime}a).  To be more precise, we also analyze the
statistics of the day-to-day differences in $\Lambda$. In Fig.
\ref{PL-wwc} we can see a clear power-law behavior emerging, again
very similar to the one we get for real markets.

In order to shed light on these findings, let us consider the
behavior of the model in the limit $\tau\to\infty$. We assume
that, in this limit, %\cite{limits},
the correlations and hence $\ket{z_t}$ reach a time independent
limit, which by Eqs. (\ref{eq1}, \ref{eq2}) is given by
\begin{equation}\label{z}
\ket{z} = \frac{1}{\hat C - \nu\epsilon}\left(\nu\ket{b} + \sigma\ket{1}\right)
\end{equation}
where $\sigma$ and
$\nu$ are fixed by the constraints $\braket{z}{1}=W$ and
$\braket{z}{r} = R$. Notice that $\ket{r}=E[\ket{\delta
x}]=\ket{b}+\epsilon\ket{z}$ needs to be determined
self-consistently. 
We find 
  $\hat C=\hat B + \Delta\ket{z}\bra{z}$ which, combined with Eq.
  (\ref{z}),
yields an equation for $\hat C$.
In order to make the analysis simpler, we assume
structure-less {\em bare} correlations $\hat B=B\hat I$. In this case $\hat C$
has $N-1$ eigenvalues equal to $B$, and one eigenvalue with
eigenvector parallel to $\ket{z}$, whose value is
\cite{2solutions}

\begin{equation}\label{solution}
\Lambda = B + \frac{\Delta W^2}{N}+\frac{N\overline{\delta
b^2}\Delta(1-\sqrt{1-a})^2}{4\epsilon}
\end{equation}
where $a=4[W(\overline{b}+\epsilon W/N)-R]/(N\overline{\delta
b^2})$ and $\overline{b}$, $\overline{\delta b^2}$ are the average
and the variance of {\em bare} returns, respectively. If $R$ and $W$ are
both proportional to $N$, then the contribution to $\Lambda$ due
to portfolio investment is also proportional to $N$. This is
indeed the order of $\Lambda$ in empirical data.
Most remarkably, Eq. (\ref{solution}) makes sense only for
$a<1/4$, i.e. for
\begin{equation}\label{Wstar}
  W<W^*=\frac{N}{2\epsilon}\left[\sqrt{\overline{b^2}+\frac{4\epsilon R}{N}}
  -\overline{b}\right].
\end{equation}
As $W\to W^*$ the solution develops a singularity with infinite
slope $\frac{\partial\Lambda}{\partial W} \to \infty$. This is
reminiscent of the divergence of susceptibility $\chi$ close to a
phase transition, signalling that the response $\delta\Lambda=\chi
\delta W$ to a small perturbation $\delta W$ diverges as $W\to
W^*$. The origin of the singularity at $W^*$ is directly related
to the impact of portfolio investment. Indeed, notice that the two
constraints are hyper-planes in the space of portfolios $\ket{z}$
when $\epsilon=0$ and always have a non-empty intersection. When
$\epsilon>0$, the constraint on return becomes an hyper-sphere,
centered in $-\ket{b}/(2\epsilon)$ and of radius
$\sqrt{\braket{b}{b}/4+\epsilon R}/\epsilon$. Hence
intersections exist only for $W<W^*$.

As anticipated, Eq. (\ref{eq1}) can be thought of as the lowest order of a
phenomenological expansion \cite{Landau}. Higher orders, e.g.
$\ket{z_{t+1}}-\ket{z_t}$, as well as terms proportional to $\ket{r_t}$ and its 
time derivatives, can be included. Likewise, one
can consider a generic matrix $\hat B$, or add several components $\ket{z^{k}_t}$ 
of portfolio investment in Eq. (\ref{eq1}), each with different parameters $R^k$, 
$W^k$ and $\Delta^k$ or acting over different time horizons $\tau_k$.
In all these case, we confirmed \cite{elsewhere} the existence of a dynamical 
instability when the 
volume of trading exceeds a critical value, as long as the time-scales ($\tau_k$) 
over which averages are taken in Eqs. (\ref{ccc},\ref{cc}) are very large. 
The analytic approach can be extended to finite $\tau$ by a
systematic $1/\tau$ expansion in the $W<W^*$ phase \cite{elsewhere}. 
This expansion describes how fluctuations in slow quantities,
such as $\ket{z_t}$ or $\hat C_t$ vanish as $\tau\to\infty$. We
find that the coefficients of the $1/\sqrt{\tau}$ expansion diverge as
$W\to W^*$, signalling that fluctuations do not vanish for
$W>W^*$. For example, we find that fluctuations  in $\Lambda$
diverge as $\delta\Lambda\sim | W^*-W|^{-1/2}$, when $W\to W^*$.
This is why higher order terms such as
$\ket{z_{t}}-\ket{z_{t-1}}$ in Eq. (\ref{eq1}) are irrelevant, in
the sense of critical phenomena, i.e. their presence does not
affect the occurrence of the phase transition.

\begin{figure}[ht]
  \centerline
      {
        \hbox
            {
            \psfig{figure=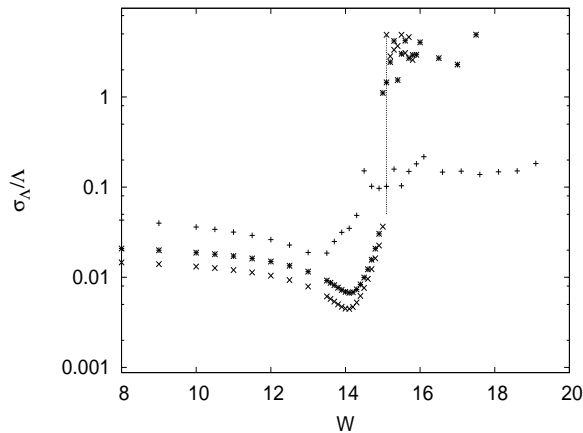,height=6cm,width=8cm}
            }
      }
      \caption{Relative fluctuation of the maximum eigenvalue
        as a function of $W$ in a simulation of the model with $N=20$,
        $\epsilon=0.1$, $R=1$, $\Delta=1$,$B=10^{-2}$,$\tau=1000$ ($+$), $\tau = 20000$
        ($\times$) and $\tau = 50000$ ($\star$). Vertical line is the theoretical
        critical value of $W$.}
      \label{fig3}
\end{figure}

Numerical simulations fully confirm these results. Fig. \ref{fig3}
reports the relative fluctuations of $\Lambda_t$ as a function of
$W$, for simulations carried out at different time scales $\tau$.
For $W<W^*$, fluctuations vanish as $\tau$ increases and
$\Lambda$ converges to the value of Eq. (\ref{solution}). For
$W>W^*$, instead, the dynamics is characterized by persistent
instabilities with fluctuations of the same order of $\Lambda$,
and it does not attain a smooth limit as $\tau\to\infty$. 
For values of $W$ smaller but close to $W^*$ the model exhibits 
strong fluctuations, precursors of the instabilities for $W>W^*$. 
It is precisely in this critical region that we recover realistic
results, such as those of Fig \ref{Lambdatime}b. Moreover, the distribution 
of returns develops a power law behavior as $W$ approaches $W^*$ (with
a cutoff which diverges as $1/\sqrt{W^*-W}$).

%{\bf Hence, the direction of $W$ is a relevant one, with the emergence of a critical point, whereas others parameters are irrelevant (in the RG sense).}

The presence of a phase transition from a stable to an unstable
state and the strong resemblance of the dynamics of the model
close to criticality with real data (see Fig. \ref{Lambdatime})
suggests that real markets might be close to the phase transition.
In order to investigate this issue systematically, we estimate the
parameters of our model from real data. In doing this we
implicitly assume that parameters $\epsilon$, $R$, $W$ etc. vary
slowly on time scales of order $\tau$. We compute the likelihood
that the particular set of time series of a given market are
produced as output of Eq. (\ref{eq1}) for a particular choice of
parameters. Next we find the parameters which maximize the
likelihood \cite{elsewhere}. As a check, the procedure was run on
synthetic data set generated by Eq. (\ref{eq1}) and it allowed us
to recover the parameters with which the data set was created. In
the inset of Fig. \ref{PL-wwc} we plot the result of such a fit
for (the assets of) four different indices in the time period
1997-2005 \cite{data}. We used $\tau=50$ and fits were taken on a
time window of $T=300$ days. Notice, in this respect, that while
in our model $\tau$ enters both in the dynamics and in the way we
take averages, in the empirical analysis it only enters in the way
we take averages, whereas we don't have access to the time scale
$\tau$ used by investors. We checked that the main results do not
depend significantly on the choice of $\tau$.  We see that fitted
parameters for real markets tend to cluster close, but below, the
transition line $W=W^*$. This is also consistent with the
similarity of the distribution of $\Lambda$ for real and synthetic
data of Fig. \ref{PL-wwc}.

%\begin{figure}[ht]
 % \centerline
  %    {
%        \hbox
%            {
%              \psfig{figure=time.eps,height=6cm,width=8cm}
%            }
%      }
%      \caption{Temporal evolution of the
%       maximum eigenvalue of the correlation matrix
%        for different values of $W$ for $N=20$.
%        $\Delta=4$, $R=20$.
%        }%%
%
%      \label{time}
%\end{figure}

Our model is very stylized and it misses many important aspects.
For example, it is undeniable that external factors and global
events have an effect on financial markets. For example, the
introduction of the Euro has a visible effect on the scatter of
points for the DAX in the inset of Fig. \ref{PL-wwc}. On the other
hand, the points relative to non-European markets in different
time windows cluster in the same region, showing that parameters
can indeed be considered as roughly constant on the time-scales
discussed here. At any rate, rather than insisting on the validity
of the model on theoretical grounds, we have shown that it
reproduces key empirical features of real financial markets. This
makes us conjecture that a sizeable contribution to the collective
behavior of markets arises from its internal dynamics and that
this is a potential cause of instability. If, following Ref. \cite{crashes},
crashes were activated events triggered by large fluctuations, the 
proximity to the instability would make the occurrence of such 
correlated fluctuations more likely, thus enhancing the likelihood
of crashes.

Our results indicate the existence of an 
additional component of risk due to the
enhanced susceptibility of the market. Such ``market impact'' risk
arises 
because investing in risk minimization %according to an optimal portfolio
strategies affects the structure of correlations with which those
strategies were computed.
This component of risk diverges as the
market approaches the critical point $W^*$, thus discouraging
further investment. This provides a simple rationale of why
markets ``self-organize'' close to the critical point. Such a
scenario is reminiscent of the picture which Minority Games
\cite{MG} provide of single asset markets as systems driven to 
a %close to 
critical state, by speculative trading \cite{GCMG}. 
In both cases, the action of traders (either to exploit predictable patterns or to
minimize risk) produces a shift in the position of the equilibrium that counteracts 
the effect of this action (by either making the market less predictable or more 
risky), as if a sort of generalized Le Chatelier's principle were at play.

%That model describes a situation where speculators are attracted by an information rich, predictable market. This, however, gets less and less predictable because of the very impact of their trading. This pushes the (Minority Game's) market close to the critical point where it becomes information-efficient. It is indeed precisely there that Minority Game models exhibit statistical properties very similar to those of real financial markets \cite{GCMG}.

%Our findings provide a further example of a case where, when the scale of human activity reaches a critical point, the collective properties of the system change dramatically. In all these cases, the assumption that the individual intervention has no impact on the aggregate -- the so-called price taking behavior here -- fails because the system reaches a point of infinite susceptibility. Such problems naturally arise where agents rely on the exploitation of a {\em public good} \cite{commons}. Their appearance in financial markets -- the prototype examples of perfect competition -- is somewhat paradoxical.


\begin{thebibliography}{200}
\bibitem{nota_ec}
The persistence of the structure of correlations \cite{mantegna,kertesz}
and their overlap with economic sectors \cite{gopicorr,states} suggests
that it can be related to economic factors which evolve on long time scales.
Financial activity is instead relevant over time scales of few days and it is
responsible for the emergence of the market mode.
\bibitem{bra-ket} $\ket{x}$
should be considered as a column vector, whereas $\bra{x}$ is a
row vector. Hence $\avg{x|y}$ is the scalar product and $\ket{x}\bra{y}$
is the direct product, i.e.
the matrix with entries $a_{i,j}=x_iy_j$.
\bibitem{noteavg} Exponential moving averages such as that used in Eqs. (\ref{ccc},\ref{cc})
are widely used in finance (see \cite{elton} p. 59). Our main results are related to the 
behavior of the model for $\tau\to\infty$ and clearly remain unchanged if one assumes 
uniform averages over a finite window $T$ (as e.g. Refs. \cite{mantegna,kertesz,gcalda,states}), and then lets $T\to\infty$.
\bibitem{2solutions} There are indeed two solutions for
  $\Lambda$.  We choose the one for which $\Lambda\to B$ as
  $\epsilon\to 0$.
\end{thebibliography}
\end{document}